\documentclass[sigconf]{acmart}

\usepackage{graphicx} 
\usepackage{float}

\usepackage{balance}
\usepackage{bm}
\usepackage{booktabs} 
\usepackage{multirow}
\usepackage{subcaption}
\usepackage{enumitem}
\usepackage{balance}
\usepackage{enumitem}
\usepackage{todonotes}
\setlist[itemize]{leftmargin=*}
\usepackage{amsmath} 
 
\usepackage{amssymb}
\usepackage{pdfrender}
\usepackage{pifont}

\usepackage[hang,flushmargin]{footmisc}

\def\vb{{\bm{b}}}

\def\vp{{\bm{p}}}
\def\vq{{\bm{q}}}
\def\vr{{\bm{r}}}
\def\vs{{\bm{s}}}

\def\vx{{\bm{x}}}

\DeclareMathAlphabet{\mathsfit}{\encodingdefault}{\sfdefault}{m}{sl}
\SetMathAlphabet{\mathsfit}{bold}{\encodingdefault}{\sfdefault}{bx}{n}

\def\sI{{\mathbb{I}}}

\def\sQ{{\mathbb{Q}}}

\usepackage{tabularx}
\usepackage{diagbox}

\AtBeginDocument{%
  }

\copyrightyear{2024} 
\acmYear{2024} 
\setcopyright{acmlicensed}\acmConference[CIKM '24]{Proceedings of the 33rd ACM International Conference on Information and Knowledge Management}{October 21--25, 2024}{Boise, ID, USA}
\acmBooktitle{Proceedings of the 33rd ACM International Conference on Information and Knowledge Management (CIKM '24), October 21--25, 2024, Boise, ID, USA}
\acmDOI{10.1145/3627673.3679849}
\acmISBN{979-8-4007-0436-9/24/10}

\begin{document}

\title{Enhancing CTR prediction in Recommendation Domain with Search Query Representation}

\author{Yuening Wang}
\authornote{Both authors contributed equally to this research.}
\affiliation{%
  \institution{Huawei Noah's Ark Lab}
  \city{Markham}
  \country{Canada}
}
\email{yuening.wang@huawei.com}

\author{Man Chen}
\authornotemark[1]
\affiliation{%
  \institution{Huawei Noah's Ark Lab}
  \city{Montreal}
  \country{Canada}
}
\email{man.chen1@huawei.com}

\author{Yaochen Hu}
\affiliation{%
  \institution{Huawei Noah's Ark Lab}
  \city{Montreal}
  \country{Canada}
}
\email{yaochen.hu@huawei.com}

\author{Wei Guo}
\affiliation{%
  \institution{Huawei Noah's Ark Lab}
  \city{Singapore}
  \country{Singapore}
}
\email{guowei67@huawei.com}

\author{Yingxue Zhang}
\affiliation{%
  \institution{Huawei Noah's Ark Lab}
  \city{Markham}
  \country{Canada}
}
\email{yingxue.zhang@huawei.com}

\author{Huifeng Guo}
\affiliation{%
  \institution{Huawei Noah's Ark Lab}
  \city{Shenzhen}
  \country{China}
}
\email{huifeng.guo@huawei.com}

\author{Yong Liu}
\affiliation{%
  \institution{Huawei Noah's Ark Lab}
  \city{Singapore}
  \country{Singapore}
}
\email{liu.yong6@huawei.com}

\author{Mark Coates}
\affiliation{%
  \institution{McGill University}
  \city{Montreal}
  \country{Canada}}
\email{mark.coates@mcgill.ca}






\begin{abstract}
Many platforms, such as e-commerce websites, offer both search and recommendation services simultaneously to better meet users' diverse needs. Recommendation services suggest items based on user preferences, while search services allow users to search for items before providing recommendations. Since users and items are often shared between the search and recommendation domains, there is a valuable opportunity to enhance the recommendation domain by leveraging user preferences extracted from the search domain. Existing approaches either overlook the shift in user intention between these domains or fail to capture the significant impact of learning from users' search queries on understanding their interests.

In this paper, we propose a framework that learns from user search query embeddings within the context of user preferences in the recommendation domain. Specifically, user search query sequences from the search domain are used to predict the items users will click at the next time point in the recommendation domain. Additionally, the relationship between queries and items is explored through contrastive learning. To address issues of data sparsity, the diffusion model is incorporated to infer positive items the user will select after searching with certain queries in a denoising manner, which is particularly effective in preventing false positives. Effectively extracting this information, the queries are integrated into click-through rate prediction in the recommendation domain. Experimental analysis demonstrates that our model outperforms state-of-the-art models in the recommendation domain.
\end{abstract}




\maketitle

\section{Introduction}\label{introduction}

The rapid expansion of online information has underscored the essential roles of recommender systems and search engines in enabling users to efficiently access relevant and appealing content. Numerous online platforms and applications, such as Amazon, Kuaishou, and YouTube, provide both recommendation and search services concurrently to adequately meet diverse user needs. Recommender systems offer suggestions to users as they browse, whereas search services allow users to actively seek out specific items. The integration of user interactions across these services on the same platforms presents a valuable opportunity to leverage behavior data from both domains, enhancing the accuracy of click-through rate (CTR) predictions in recommender systems.

Several studies have explored joint learning frameworks across recommendation and search domains. IV4REC+ \cite{IV4REC} and SESRec \cite{SESRec} incorporate search data to better understand user behavior in the recommendation domain. USER \cite{USER} and SRJGraph \cite{SRJGraph} integrate both domains by taking the recommendation behavior as a special case of the search behavior with zero queries. 

Despite these advances, several challenges remain unaddressed. Existing approaches often neglect the significant role of user-generated queries in search services, which not only reflect explicit user interests but also influence subsequent user behaviors by exposing different sets of items. For example, SESRec \cite{SESRec} considers search queries as mere features, utilizing similar embedding techniques without introducing specialized mechanisms that can capture the rich information queries convey. Additionally, these models assume a uniformity of user interests across domains, disregarding potential shifts in user intent between browsing and specific searches. This assumption overlooks scenarios where a user's preferences might differ based on the context of their interactions. For example, a user who prefers pink might choose a pink umbrella while just browsing on the e-commerce platform. On the other hand, this user could select a black umbrella if he/she has searched for an ``umbrella'' because he/she is in more urgent need of an umbrella and the black one can be shipped more quickly. 

To confirm our conjecture, we conducted an analysis based on KuaiSAR-small to determine if there is indeed an interest shift between the recommendation and search domains. As shown in figure \ref{js_hist}
and section \ref{data_analysis_1}, when we compare the selected users' categorical distribution of clicked items, with JS divergence value $\geq 0.5$, the interest shift is observed. Hence, handling potential shifts in user preferences between the search and recommendation domains is essential for effectively transitioning preferences. Queries within the search domain are often directly linked to specific items or provide descriptive text about these items. A deeper understanding of the structural relationships between queries and items could lead to information-enriched query representations. Given that items overlap across domains, such insights gained from query representations can be invaluable in the recommendation domain. Previous studies, such as SESRec \cite{SESRec}, have also explored query-item alignment. However, our data analysis reveals a significant challenge: a large number of users do not actively click on items after performing searches (e.g., one-third of users in the Kuaishou-small dataset). Simply disregarding these non-click instances, as is done in other studies, prevents us from fully capturing the complexity of query-item relationships. This suggests a need for a more nuanced approach that accounts for queries with no corresponding clicked item records.
 
To address the aforementioned problem, we design a search-enhanced framework for CTR prediction named {\em QueryRec}. This framework aims to utilize the search domain to enhance prediction accuracy in the recommendation domain by integrating refined query representations. Specifically, the query history is explicitly added to the recommendation domain as an augmented feature. To fully exploit the interaction history in the search domain, we design two auxiliary training losses to enhance the learned query embeddings: next-item prediction to align the query embedding with the user's interests in the recommender domain and query-item constructive learning to implicitly align the query embedding with their correlated items. Furthermore, we adopt a diffusion module to alleviate the issue that a significant amount of queries do not have positively interacted items. 

In the next-item prediction module, we align the representation of the query list and the next clicked item in the recommendation domain. We adopt the self-attention sequential encoder \cite{Kang2018SelfAttentiveSR} to aggregate the embeddings of all the queries from the query list and push them close to the next clicked items. 
This approach directly aligns the query representation with user interests in the recommendation domain, effectively addressing the issue of interest shifts. 

We implement a contrastive loss to capture query-item relations in the search domain and align the query embeddings with the clicked item embeddings under that search query. 
To alleviate the issue of sparse positively interacted items for queries (even no positively interacted items for a significant amount of queries), we train a diffusion model to predict how likely the user will click an item given the positive and negative lists of items. We select the top$K$ items (those assigned the highest probability by the generative model) to augment the positive list items. 
This allows the diffusion model to effectively perform a form of denoising when generating the positive items, which can effectively avoid false positives \cite{ho2020denoising}. Besides, this also serves as an imputation mechanism for the positive list of items for those queries without an empty one.  

In summary, we make the following contributions:
\begin{itemize}
\item This framework represents the first known effort to focus on the representation learning of queries, which are crucial for capturing user preferences within the search domain. This approach enables a more effective transfer of knowledge to the recommendation domain.
\item We introduce a next-item prediction module that integrates information from both the search and recommendation domains, effectively bridging the gap in user interests between these two domains. This integration ensures that the information transferred to the recommendation domain is more accurately aligned with target performance expectations.
\item We propose a constrictive learning module with the samples in the search domain to better align the query embedding and item embedding. Furthermore, We design and train a diffusion model to generate positive item suggestions based on the clicked and non-clicked item lists under specific queries, which better extract training labels from the non-clicked item list. 

\end{itemize}

\section{Related work}
\subsection{Recommendation systems}
Numerous cross-domain recommendation systems have been devised to harness shared knowledge across different domains, thereby enhancing the performance within target domains. Leveraging common users or items as conduits for knowledge transfer, certain models employ these shared entities as bridges. For instance, in \cite{Rafailidis2016TopNRV, Rafailidis2017} matrix factorization is applied to user-item interactions within each domain, using overlapping user or item information as constraints to guide the factorization process. The approaches in \cite{DTCDR2019, CSRZhu2018} combine representations of overlapping users or items to transfer information, whereas mapping functions are trained to project representations between overlapping users or items between the two domains in \cite{EMCDR2017, Zhu2022PTUPCDR}. Graph-based methods such as the technique of \cite{HEROGRAPH2022} build user-item interaction graphs using overlapping user/item nodes to link the different domains. Non-linear and high-order cross-domain knowledge is captured through message passing on the graph. Some other models, instead of directly associating users and items between different domains, assume other content to be related, such as rating patterns \cite{CDRSCTZhang2019}, or tag/review information \cite{CDIE-C, CFAA}. The typical process involves mapping the user-item interaction matrix to the same latent space, which captures a common pattern shared by the different domains, enabling knowledge transfer. However, these methods often do not fully utilize the rich data available from user behaviors or contextual features, which are frequently highly informative.

\subsection{Cross-domain recommendation systems}
Some existing methods focus on click-through rate (CTR) prediction problems in different domains. PLE \cite{Tang2020PLE} improves MMoE \cite{MMoE} by explicitly separating domain-shared and domain-specific experts to avoid negative transfers. 3MN \cite{3MN} adds three meta-networks to PLE, so that the embedding layer is capable of learning the domain-related knowledge explicitly. STAR \cite{STAR} facilitates effective information transfer between different domains by combining the weights of a shared fully connected network (FCN) and a domain-specific FCN. Shen et al. develop SAR-net \cite{Shen2021sarnet}, which applies an attention mechanism to merge domain-specific and domain-shared information. MiNet \cite{Ouyang2020MiNet} learns users' long-term domain-shared interests and short-term domain-specific interests. CCTL \cite{CCTL} applies different sample importance weights when transferring information from the source domain to the target domain in order to only transfer useful knowledge and avoid conflicting information. Liu et al. \cite{ALH} introduce a novel adaptive loss to harmonize training across multiple domains. Although these methods have shown promising performance, they are suboptimal for our problem. First, as these models are designed for general cross-domain systems, they do not explicitly extract users' interests from search queries. Search queries are strong signals that reflect users' preferences. Most of the models do not consider how unique features within the different domains can be more effectively processed or learned via the application of different mechanisms.

\subsection{Joint learning of search and recommendation}
There are also existing works that focus on joint learning of search and recommendation. Zamini et al. \cite{Zamani2018JointMA, Zamani2020} train models in both the search and the recommendation domain with a joint loss and shared item information. IV4REC+ \cite{IV4REC} enhances recommendation with search data through a causal learning framework. The search queries are used as treatment variables to decompose item embedding vectors into a causal part and a non-causal part. The causal part better reflects why a user prefers an item. By doing so, they implicitly assume that the user intentions in the search domain could represent that of the recommendation domain without considering the interest shift across the two domains. SESRec \cite{SESRec} leverages users' search interests for the recommendation domain by disentangling similar and dissimilar representations within the search and recommendation behavior. However, the dissimilar and similar representations are then concatenated together for prediction, not further distinguishing the transferable and non-transferable parts between the recommendation and the search domains. USER \cite{USER} integrates users' behavior in search and recommendation into a heterogeneous behavior sequence by treating recommendation as search with an EMPTY query. SRJGraph \cite{SRJGraph} builds a unified graph between the search and recommendation domains, using a query as an edge. A specific ``zero''-type query is used as the edge for the recommendation domain. These two works overlook that the item exposure mechanism is different in two domains. For recommendation, the items exposed depend on the put feature information, whether the item is exposed in the search domain is mainly determined by the queries searched. Therefore, simply considering the recommendation behavior as a special search behavior is not optimal. 

In contrast, our model matches the interest of the two domains by predicting items clicked at the next time point in the recommendation domain with historical query search history in the search domain. Besides, the model addresses the importance of the query search information and the query-item relationship with an additional diffusion-augmented contrastive learning module.

\section{Preliminary material}
\subsection{Problem formulation}
\begin{table}[t]
\centering
\caption{Important Notations}
\label{tab:notation1}
\resizebox{\linewidth}{!}
{\begin{tabular}{c|cc}
		\toprule
\textbf{Symbol}&&\textbf{Definition}\\\hline
 $i \in \mathbb{I}$&& an item \\
            
           $\mathbb{I}$&& the set of items \\
           
            $u \in \mathbb{U}$&& a user \\
            
           $\mathbb{U}$&& the set of users \\
           
           $t$ && the time of the corresponding interaction\\

           $\mathbf{b^u}_t$ && the user $u$'s historical clicked items before time $t$\\
           
$q \in \mathbb{Q}$&& a query\\
           
$\mathbb{Q}$ &&the set of queries \\
           
             $\mathbf{q^u}_t$ && the user $u$'s historical queries  before time $t$\\   
             $y \in \{0, 1\}$ && the label indicating whether the click action happened\\   
\bottomrule
  \end{tabular} }
\end{table}


Consider a recommendation domain and a search domain that share the same set of users $\mathbb{U}$ and items $\mathbb{I}$. Each user $u\in \mathbb{U}$ is associated with a vector of categorical features $\vx_u$, and each item has a vector of categorical features $\vx_i$. 

In the recommendation domain, we have a collection of user-item interactions. Each interaction is a 4-tuple $(u, i, t, y)$, where $u$ and $i$ are the user and item index, $t$ is the time when the interaction occurred, and $y\in\{0, 1\}$ indicates whether the user $u$ clicked the item $i$. We associated with each user $u$ a list $\vb^t_u$ containing all the items that the user $u$ clicked before time $t$. 

The search domain has a set of queries $\sQ$. Each interaction is a 5-tuple $(q, u, i, t, y)$, where $q\in\sQ$ is the query, $u\in \mathbb{U}$ is the user, $i\in \mathbb{I}$ is the item, $t$ is the time of the interaction, and $y\in\{0, 1\}$ indicates whether the user $u$ clicked the item $i$ after searching with query $q$. A query is a string that a user inputs to the search domain platform to find some items. Associated with each user $u$ is a list $\vq^t_u$ containing all the queries conducted by the user $u$ before time $t$. 

We aim to utilize the historical interactions from both the recommendation and search domains to predict $\Tilde{y}$, the probability of a user $u$ clicking the item $i$, i.e., the click-through rate (CTR). Specifically, to make the most of the information from both domains, the model takes input 
$(u, i, t, \vx_u, \vx_i, \vb^t_u, \vq^t_u)$,
and predict $\Tilde{y}$. We call $\vb^t_u$ and $\vq^t_u$ the user behavior features. Compared with the single recommendation domain setting, the user query list $\vq^t_u$ in the task input is explicitly augmented from the search domain. 

\subsection{Backbone models}
\label{sec:backbone}
There has been enormous work on CTR prediction \cite{DIN, DIEN, DSIN, SIM, DMIN}. We call them the backbone models, which can be directly applied to the input in the task input. The typical designs utilize an 
embedding layer to transfer all the input features into embeddings, followed by various neural network structures to extract the deep correlation between the features. 
Without loss of generality, we take DIN \cite{DIN2018} as an illustration example in this work.
In DIN, for each user behavior feature, it utilizes a target-aware adaptive pooling to aggregate the behavior embeddings of the items to learn enhanced user interest representations. 
Those enhanced user interest representations are concatenated with the embedding of other features. Finally, an MLP is adopted to generate the final prediction.

\textbf{Embedding Layer.}
The embedding layer is used to convert the sparse, high-dimensional input into dense, low-dimensional embedding vectors via lookup to an embedding table. 
Specifically, we denote $\boldsymbol{e}_u$ as the user embedding and $\boldsymbol{e}_i$ as the item embedding. Each element in $\vx_u$ is mapped into an embedding, and we denote the concatenated embedding as $\boldsymbol{e}_{\vx_u}$. Similarly, we define $\boldsymbol{e}_{\vx_i}$ for $\vx_i$. For $\forall j\in \vb_u^t$, we have $\boldsymbol{e}_j$ that share the same set of embeddings with the item embeddings. For $\forall q\in \vq_u^t$, we have $\boldsymbol{e}_q$. 

\textbf{Prediction with Adaptive Pooling.} Adaptive pooling is adopted to model user interest from behavioral features. Specifically, for $\vb_u^t$, and the target item $i$, we define the user's interest representation from $\vb_u^t$ as
\begin{align}
      \boldsymbol{\nu}_{\vb_u^t} = \sum_{j\in\vb_u^t} a(\boldsymbol{e}_j, \boldsymbol{e}_{i})\boldsymbol{e}_j,
\end{align}
where $\boldsymbol{e}_{i}$ is the embedding of target item $i$, $a(\cdot, \cdot)$ represents some feed-forward neural network model for the adaptive weights. Similarly, for $\vq_u^t$, and the target item $i$, we define the user's interest representation from $\vq_u^t$ as
\begin{align}
      \boldsymbol{\nu}_{\vq_u^t} = \sum_{j\in\vq_u^t} a(\boldsymbol{e}_j, \boldsymbol{e}_{i})\boldsymbol{e}_j.
\end{align}
Then, we concatenate all the embeddings
\begin{align}
        \boldsymbol{e}_{\text{in}} = (\boldsymbol{e}_u\| \boldsymbol{e}_i\| \boldsymbol{e}_{\vx_u}\| \boldsymbol{e}_{\vx_i}\| \boldsymbol{\nu}_{\vb_u^t} \| \boldsymbol{\nu}_{\vq_u^t})
\end{align}
$\boldsymbol{e}_{\text{in}}$ is then fed into an MLP to get the final prediction $\tilde{y}$, i.e., 
\begin{align}
    \tilde{y} = \text{MLP}(\boldsymbol{e}_{\text{in}}).
\end{align}

\textbf{Loss function.} Cross entropy loss is employed. For each sample $(u, i, t, y)$ in the training set in the recommendation domain, we have
\begin{align}
    \mathcal{L}_1 = -y\cdot\text{log}\tilde{y} - (1-y)\cdot\text{log}(1-\tilde{y}).
\end{align}

\subsection{Diffusion model for recommender systems}\label{diffusion processes}
The diffusion model is a powerful generative model \cite{wang2023diffusion} and \cite{ho2020denoising} adapt it for recommender systems. 
Specifically, let $\vx_0\in \mathcal{R}^{|\sI|}$ denote the interaction vector for a user, where the $i^{th}$ element is $1$ if the user has interacted with item $i$; otherwise it is $0$. 
Based on this vector, the diffusion model generates $\tilde{\vx}_0$ to predict whether the user will interact with all the other items. 
\paragraph{Forward process}
From the initial state $\vx_0$, the iterative forward process can be parameterized as:
\begin{equation}
q\left(\mathbf{x}_t \mid \mathbf{x}_{t-1}\right)=\mathcal{N}\left(\mathbf{x}_t ; \sqrt{1-\beta_t} \mathbf{x}_{t-1}, \beta_t I\right),
\end{equation}
where $\beta_t \in(0,1)$ controls the Gaussian noise scales added at each step $t$. As we know, by adding two independent Gaussian noises together, we can get a new Gaussian noise. By reparameterization trick, we can directly obtain $\mathbf{x}_t$ from $\mathbf{x}_0$:
\begin{equation}
q\left(\mathbf{x}_t \mid \mathbf{x}_0\right)=\mathcal{N}\left(\mathbf{x}_t ; \sqrt{\bar{\alpha}} \mathbf{x}_0,\left(1-\bar{\alpha}_t\right) I\right),
\end{equation}
where $\alpha_t=1-\beta_t, \bar{\alpha}_t=\prod_{t^{\prime}=1}^t \alpha_{t^{\prime}}$, and then
\begin{equation}
 \label{fwd1}   
\mathbf{x}_t=\sqrt{\bar{\alpha}_t} \mathbf{x}_0+\sqrt{1-\bar{\alpha}_t} \boldsymbol{\epsilon},
\end{equation}
can be reparameterized with $\boldsymbol{\epsilon} \sim \mathcal{N}(\mathbf{0}, I)$. The forward process is defined as Markov chain \cite{ho2020denoising}. 

\paragraph{Reversed process}
  \cite{ho2020denoising} defines the joint distribution $p_\theta\left(\mathbf{x}_{0: T}\right)$ as the reverse process. It is regarded as a Markov chain with learned Gaussian transitions starting at 
\begin{equation}
\label{rev1} 
\begin{aligned}   
& p\left(\mathbf{x}_T\right)=\mathcal{N}\left(\mathbf{x}_T ; \mathbf{0}, \mathbf{I}\right), p_\theta\left(\mathbf{x}_{0: T}\right):=p\left(\mathbf{x}_T\right) \prod_{t=1}^T p_\theta\left(\mathbf{x}_{t-1} \mid \mathbf{x}_t\right),\\
&  \quad p_\theta\left(\mathbf{x}_{t-1} \mid \mathbf{x}_t\right):=\mathcal{N}\left(\mathbf{x}_{t-1} ; \boldsymbol{\mu}_\theta\left(\mathbf{x}_t, t\right), \boldsymbol{\Sigma}_\theta\left(\mathbf{x}_t, t\right)\right),
\end{aligned}   
\end{equation}
where $\mu_\theta\left(\mathbf{x}_t, t\right)$ and $\Sigma_\theta\left(\mathbf{x}_t, t\right)$ are the mean and covariance of the Gaussian distribution predicted by a neural network with parameters $\theta$. According to Baye's rule and the property of the Markov chain, $q\left(\vx_{t-1} \mid \vx_t, \vx_0\right)$ can be rewritten as the following:

\begin{equation}
\label{rev1} 
\begin{aligned}   
& q\left(\vx_{t-1} \mid \vx_t, \vx_0\right) \propto \mathcal{N}\left(\vx_{t-1} ; \tilde{\boldsymbol{\mu}}\left(\vx_t, \vx_0, t\right), \sigma^2(t) I\right), \text { where } \\
& \left\{\begin{array}{l}
\tilde{\boldsymbol{\mu}}\left(\boldsymbol{x}_t, \boldsymbol{x}_0, t\right)=\frac{\sqrt{\alpha_t}\left(1-\bar{\alpha}_{t-1}\right)}{1-\bar{\alpha}_t} \mathbf{x}_t+\frac{\sqrt{\bar{\alpha}_{t-1}}\left(1-\alpha_t\right)}{1-\bar{\alpha}_t} \mathbf{x}_0, \\
\sigma^2(t)=\frac{\left(1-\alpha_t\right)\left(1-\bar{\alpha}_{t-1}\right)}{1-\bar{\alpha}_t} .
\end{array}\right.
\end{aligned}   
\end{equation}
$\tilde{\boldsymbol{\mu}}\left(x_t, x_0, t\right)$ and $\sigma^2(t) \boldsymbol{I}$ are the mean and covariance of $q\left(\mathbf{x}_{t-1} \mid \mathbf{x}_t, \mathbf{x}_0\right)$
\paragraph{Optimization}
DDPM \cite{ho2020denoising} is optimized by maximizing the Evidence Lower Bound (ELBO) of the likelihood of observed input data $\vx_0$ :
\begin{equation}
\label{opt1} 
\begin{aligned}
& \log p\left(\vx_0\right)=\log \int p\left(\vx_{0: T}\right) \mathrm{d} \vx_{1: T} \\
& =\log \mathbb{E}_{q\left(\vx_{1: T} \mid \vx_0\right)}\left[\frac{p\left(\vx_{0: T}\right)}{q\left(\vx_{1: T} \mid \vx_0\right)}\right] \\
& \geq \underbrace{\mathbb{E}_{\boldsymbol{q}\left(\boldsymbol{x}_1 \mid \boldsymbol{x}_0\right)}\left[\log p_\theta\left(\boldsymbol{x}_0 \mid \boldsymbol{x}_1\right)\right]}_{\text {(reconstruction term) }}-\underbrace{D_{\mathrm{KL}}\left(q\left(\boldsymbol{x}_T \mid \boldsymbol{x}_0\right) \| p\left(\boldsymbol{x}_T\right)\right)}_{\text {(prior matching term) }} \\
&
-\sum_{t=2}^T \underbrace{\mathbb{E}_{q\left(\boldsymbol{x}_t \mid \boldsymbol{x}_0\right)}\left[D_{\mathrm{KL}}\left(q\left(\boldsymbol{x}_{t-1} \mid \boldsymbol{x}_t, \boldsymbol{x}_0\right) \| p_\theta\left(\boldsymbol{x}_{t-1} \mid \boldsymbol{x}_t\right)\right)\right]}_{\text {(denoising matching term) }}.
\end{aligned}
\end{equation}

The denoising matching term $\mathcal{L}_t$ at step $t$ can be calculated by

\begin{equation}
\begin{aligned}
\mathcal{L}_t & \triangleq \mathbb{E}_{q\left(\mathbf{x}_t \mid x_0\right)}\left[D_{\mathrm{KL}}\left(q\left(\mathbf{x}_{t-1} \mid \mathbf{x}_t, \mathbf{x}_0\right) \| p_\theta\left(\mathbf{x}_{t-1} \mid \mathbf{x}_t\right)\right)\right] \\
& =\mathbb{E}_{q\left(\mathbf{x}_t \mid \mathbf{x}_0\right)}\left[\frac{1}{2 \sigma^2(t)}\left[\left\|\mu_\theta\left(\mathbf{x}_t, t\right)-\tilde{\mu}\left(\mathbf{x}_t, \mathbf{x}_0, t\right)\right\|_2^2\right]\right],
\end{aligned}
\end{equation}

The simplified $\mathcal{L}_t$ is the training loss \cite{ho2020denoising, wang2023diffusion, luo2022understanding}. 
\section{Methodology}


\begin{figure*}
    \centering
    \includegraphics[width=\linewidth]{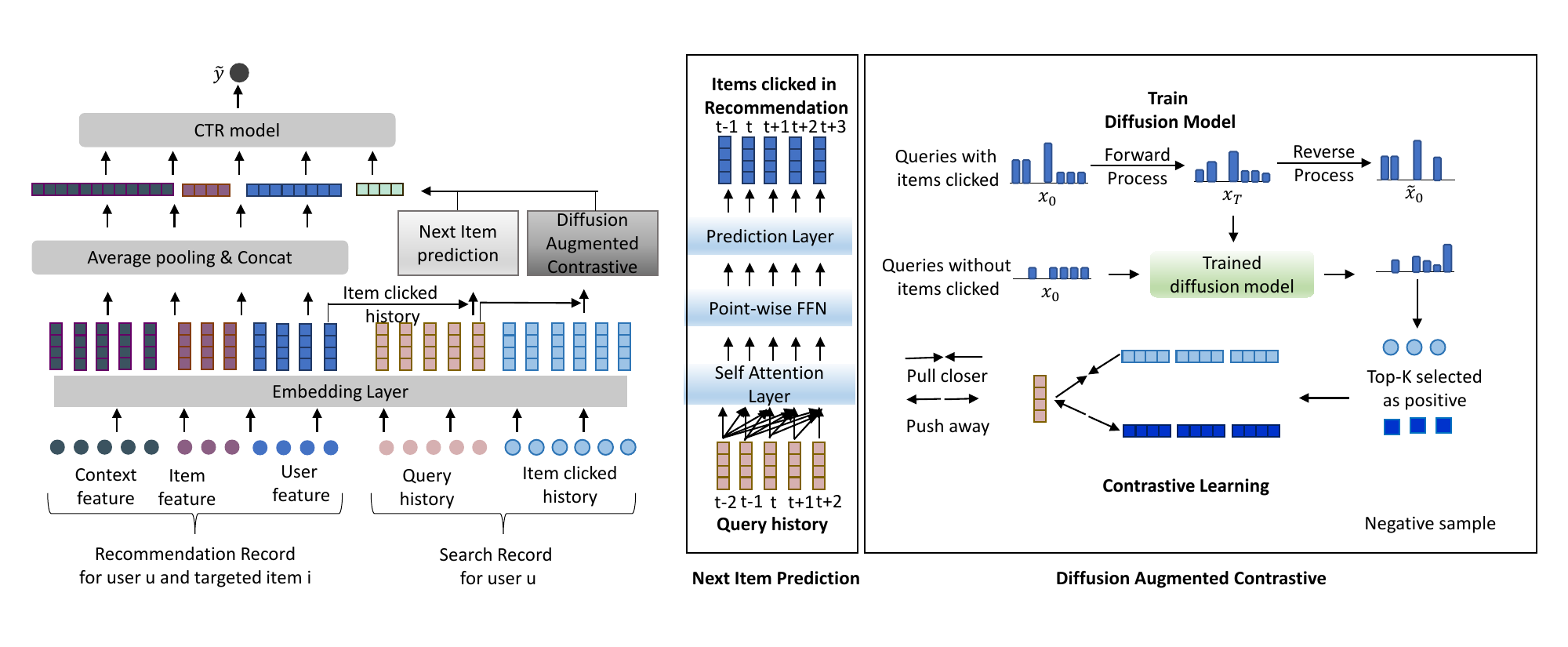}
    \caption{\textbf{The overall framework of our proposed QueryRec model}. Query representation from the search domain is fully learned by next-item prediction and diffusion-augmented contrastive learning modules. This refined representation is combined with other features from the recommendation domain and used in a CTR prediction model. For the next item prediction, the query sequences in chronological order up to time point $t$ are encoded with the self-attention layer and pointwise feed-forward network (FFN), and then predict the item the user will click in the recommendation domain at time point $t+1$. In diffusion-augmented contrastive learning, the diffusion model is first trained with records with items clicked after searching with certain queries. $x_0$ are probabilities of items that the users would click after a certain query. Considering the change the users could still be interested in the items that have not been exposed, a value from $[-1, 1]$ is also assigned to non-exposed items. After the reverse process of the diffusion model, we get $\Tilde{x}_0$, which is the original item clicked history of in the search domain. The trained diffusion model is used to learn the item distributions for queries without corresponding clicked items. The top-K items are treated as positive samples for contrastive learning, which aims to capture query-item relationships.}
    \label{fig:overview}
\end{figure*}
Any backbone model can fit the input in the task input into a prediction of CTR. However, the vanilla training method for the backbone models is not designed to extract information from the search domain and is prone to sub-optimal performance. 
First, the query list from the search domain is usually not aligned with the user's interest in the recommendation domain.
Without a proper mapping, the user query list can provide little information to the recommendation domain.
Second, there are rich historical interactions in the search domain. Without a proper mechanism, the backbone models cannot learn such information.

With the SOTA structure for CTR models, one critical issue is learning informative query embeddings that incorporate the interaction history from the search domain and utilize the correlation of recommendation and search domain. Specifically, we design the next item prediction loss to directly align aggregated embedding from the query list into the user's interest in the recommendation domain. Besides, we propose a contrastive loss to inject the relations of the embeddings between queries and items. We adopt the diffusion model to enhance the query embedding for those with few or no positive item interactions by utilizing the lists of non-clicked items. 

\subsection{Interest alignment with next item prediction}
The goal of incorporating query embedding is to help better understand users' preferences in the recommendation domain. Therefore, it is expected that the query embedding can be more informative in a way that better matches users' interest in the recommendation domain. To achieve the goal, we map interaction records of recommendation and search domains in chronological order. Then, we orient learning of query representation to directly predict the items the user intends to click at the next time point in the recommendation domain. 

In the recommendation domain, from all the interaction history $(u, i, y, t)$, we can construct two lists of items: $\vr^{+}_{u,t}$ and $\vr^{-}_{u,t}$, which are the list of items with $y=1$ and $y=0$ for user $u$ after time $t$ in chronological order. 
And similarly, a list of search sequence $\vq_t$ and interacted items in the search domain $\vs_t$. Given a training sample $(u, i, y, t)$ in the recommendation domain, we can collect the corresponding $\vr^{+}_{u,t}$, $\vr^{-}_{u,t}$ and $\vq^t_u$. The query list $\vq^t_u$ is encoded by self-attention sequential (SAS) \cite{Kang2018SelfAttentiveSR} encoder into $d$-dimensional embedding $\boldsymbol{e}_{\vq^t_u}$, which captures long-term semantics. 

On the other hand, we can get the $d$-dimensional item embeddings for the items from both $\vr^{+}_{u,t}$ and $\vr^{-}_{u,t}$ with the same embedding mechanism in Sec.~\ref{sec:backbone}. To make the query embedding reflect users' preferences in the recommendation domain better, we drive the embedding $\boldsymbol{e}_{\vq^t_u}$ to be closer to the embeddings from $\vr^{+}_{u,t}$ than those from $\vr^{-}_{u,t}$. Specifically, we define 

\begin{align}
        \mathcal{L}_2^\prime = - \frac{1}{|\vr^{+}_{u,t}|}\sum_{j\in\vr^{+}_{u,t}}\text{log}\langle \boldsymbol{e}_{\vq^t_u}, \boldsymbol{e}_j \rangle + \frac{1}{|\vr^{-}_{u,t}|}\sum_{k\in\vr^{-}_{u,t}}\text{log}(1-\langle \boldsymbol{e}_{\vq^t_u}, \boldsymbol{e}_k \rangle),
\end{align}
where $\langle\cdot, \cdot\rangle$ denotes the dot product. For efficient implementation, we sample $n_\text{pos}$ items from $\vr^{+}_{u,t}$, and denote the sampled list as ${\vr^{+}_{u,t}}^\prime$. Similar we sample $n_\text{neg}$ items from $\vr^{-}_{u,t}$ and denote it as ${\vr^{-}_{u,t}}^\prime$. We define the loss as
\begin{align}
        \mathcal{L}_2 = - \frac{1}{n_\text{pos}}\sum_{j\in{\vr^{+}_{u,t}}^\prime}\text{log}\langle \boldsymbol{e}_{\vq^t_u}, \boldsymbol{e}_j \rangle + \frac{1}{n_\text{neg}}\sum_{k\in{\vr^{-}_{u,t}}^\prime}\text{log}(1-\langle \boldsymbol{e}_{\vq^t_u}, \boldsymbol{e}_k \rangle).
\end{align}

\subsection{Query-item contrastive learning with diffusion data augmentation}
Given the user's intense activities in the search domain, it is worthwhile to capture more transferable knowledge from the search domain and thus make query representation more useful in item recommendation. One important piece of information is the relationship between the queries and the items. The probability of items being selected after certain search behaviors mostly depends on the queries searched.
The relationship is extracted with the queries and their corresponding clicked items. Inspired by \cite{Li2021AlignBF}, we leverage contrastive loss to learn similarities and dissimilarities between searched query items. 

\textbf{Contrastive learning}. From the interaction history from the search domain, for each query $q$, we collect the set of positive items for query $q$ and denote it as $\mathcal{I}^{+}_q$, where for each item $i\in\mathcal{I}^{+}_q$, at least one user clicked it after she searched with $q$ and item $i$ was exposed to her. For each sample $(u, i, t, y)$ in the recommendation domain, denote $q^t$ as the last query User $u$ has before $t$. Then, the contrastive loss at $(u, i, t, y)$ is defined as
\begin{align}
    \mathcal{L}_3 = \frac{1}{|\mathcal{I}_q^{+}|}\sum_{i \in \mathcal{I}_q^{+}}\text{log}\frac{\text{exp}(s(\boldsymbol{e}_{q^t},\boldsymbol{e}_{i})/\beta)}{\sum_{j \in \mathbb{I}\setminus\mathcal{I}_q^{+}}\text{exp}(s(\boldsymbol{e}_{q^t},\boldsymbol{e}_{j})/\beta)}\label{eq:constrastive}
\end{align}
where $s(\boldsymbol{e}_i,\boldsymbol{e}_j)=\text{tanh}(\boldsymbol{e}_{i}^T\boldsymbol{W}\boldsymbol{e}_{j})$. $\boldsymbol{W} \in \mathbb{R}^{d \times d}$ is the learnable transformation matrix between queries and items and $\mathbb{I}$ is the item set.

\textbf{Diffusion data augmentation.}\label{augmentation} One common problem for the search datasets is the sparse record of users' click behavior after searching certain queries, and a significant proportion of queries do not have positively interacted items, namely, the $\mathcal{I}_q^{+}$ is empty for a significant amount of query $q$. The query with empty $\mathcal{I}_q^{+}$ will thus have no labels in the contrastive loss. However, those query has the list of non-clicked items $\mathcal{I}_q^{-}$, where for each item $i\in\mathcal{I}^{+}_q$, no user clicked it after she searched with $q$ and item $i$ was exposed to her. We could utilize $\mathcal{I}_q^{-}$ for the contrastive signal. In light of the success of diffusion models in image generation and Top-K recommendation \cite{wang2023diffusion, ho2020denoising}, we take advantage of the diffusion module to augment information of items clicked given queries searched. We aim to enhance $\mathcal{I}_q^{+}$ with $\mathcal{I}_q^{+}$ and $\mathcal{I}_q^{-}$ by a diffusion model. 

Specifically, we take $\vx_q\in\mathcal{R}^{|\sI|}$ as the query-item interaction vector, where the $i^{th}$ element represents how likely a user will click $i$ after search with query $q$. For each query $q$, the items are divided into three groups, i.e., $\mathcal{I}_q^{+}$, $\mathcal{I}_q^{-}$, and $\sI\setminus(\mathcal{I}_q^{+}\cup\mathcal{I}_q^{-})$. For all the positions with indexes from $\mathcal{I}_q^{+}$ in $\vx_q$, we assign value $r_p\in[0, 1]$. Similarly, for all the positions with indexes from $\mathcal{I}_q^{-}$ and $\sI\setminus(\mathcal{I}_q^{+}\cup\mathcal{I}_q^{-})$ in $\vx_q$, we assign value $r_n \in[-1, 0) $ and $r_0=0 $. The $r_p$, $r_n$ and $r_0$ are hyper-parameters. 

The diffusion model is trained by the same method introduced in Sec.~\ref{diffusion processes} after we assign $\vx_0=\vx_q$. Furthermore, to learn a better model for sparse $\mathcal{I}_q^{+}$, we do some random masking to intentionally remove some items for $\vx_0$ in the forward process but keep the original $\vx_q$ as the $\vx_0$ for the reconstruction term in (\ref{opt1}).

After we trained a diffusion model, for each query $q$, we generate $\tilde{\vx}_q$. We take the index with top$K$ values in $\tilde{\vx}_q$ as the enhanced $\mathcal{I}_q^{+}$ for the contrastive loss in (\ref{eq:constrastive}).

\subsection{Training}
With all the components, we train the model with the loss
\begin{align}
    \mathcal{L} = \mathcal{L}_1 + \lambda_2\mathcal{L}_2 + \lambda_3\mathcal{L}_3,
\end{align}
where $\lambda_2$ and $\lambda_3$ are the hyper-parameters. 
\section{EXPERIMENTS}\label{experiment}
\subsection{Dataset}
\textbf{KuaiSAR} \cite{sun2023kuaisar} datasets include two versions: \textbf{KuaiSAR-large} and \textbf{KuaiSAR-small}. \textbf{KuaiSAR-large} contains data from 2023/5/22 to 2023/6/10, while \textbf{KuaiSAR-small} consists of data from 2023/5/22 to 2023/5/31. Unlike other commonly used datasets, KuaiSAR datasets are collected from a real-world platform. Due to the different selection periods, as indicated by the statistics in Table \ref{data_stat}, \textbf{KuaiSAR-large} and \textbf{KuaiSAR-small} can be considered two datasets with varying degrees of actions and correlations across two domains. It is important to note that a query is a specific list of words, with each word hashed into integers for anonymization. The query set comprises all queries from selected users, and the corresponding words form the word set.

We tested QueryRec with an industrial dataset, which includes two sub-datasets collected from two separate services: search and recommendation. The users and items of these two services are overlapped. This industrial dataset contains more than 200,000,000 records and 1,000,0000 different queries, along with other information-enriched feature fields. Besides, Only DIN, DCN, and QueryRec are tested with this industrial dataset to save extra costs.
            

           
  
\begin{table}[t]
           \caption{Statistics of public datasets.}
\label{data_stat} 
	\resizebox{\linewidth}{!}{\begin{tabular}{c|cccccc}
		\toprule
\textbf{Datasets}&\textbf{Domains}&\textbf{\#users}&\textbf{\#items} &\textbf{\#actions} &\textbf{\# queries}&\textbf{\# words} \\\hline
            
            \multirow{2}*{\textbf{KuaiSAR-}}&{Search} &2,5877&20,12476&317,1231&17,5849&8,0094 \\
            \multirow{2}*{\textbf{small}}&{Recommend} & 2,5877&228,1034& 749,3101& &\\
        ~&{Overlap} & 2,5877& 18,1392& & &\\

           \hline
            \multirow{2}*{\textbf{KuaiSAR-}}&{Search} &2,5705& 30,26189&505,9169 &28,1957 &10,7499 \\
            \multirow{2}*{\textbf{large}}&{Recommend} & 2,5705& 40,33453&1453,0179 &  & \\
           ~&{Overlap} & 2,5705& 97981& & &\\
           
\bottomrule
  
	\end{tabular}
}
\end{table}
\begin{table}[h]
   \caption{Calculated correlations between the recommendation and the search domain. The definitions of $R^{1}_{corr}$ and $R^{2}_{corr}$ are in \ref{domain_corr}.}

    \resizebox{6cm}{!}{\begin{tabular}{c|cc}
    \toprule
         Datasets& KuaiSAR-small & KuaiSAR-large  \\
         \midrule
          \bf{$R^{1}_{corr}$} &1.5134 &1.18015\\
          \cline{1-3}
          \bf{$R^{2}_{corr}$} &1.50059 & 1.2198\\
    \bottomrule   
    \end{tabular}}
    \label{tab:correlation_R}
\end{table}
\subsection{Data analysis}
\subsubsection{Empirical user 
 interest distributions in the search and the recommendation domains} \label{data_analysis_1}
 Suppose there are total $K$ categories of all items, then 
we model a user $u$ interest distribution by the frequency of categories clicked by $u$. Specifically, we denote $\mathcal{C}_u^{src}$ as the clicked items in the search domain, and $\mathcal{C}_u^{rec}$ as the clicked items in the recommendation domain. Taking the search domain as an example, then the interest distribution of $u$ over $K$ categories is:

\begin{equation}
    p_{u}^{src}(k) = \frac{\sum_{i \in \mathcal{C}_u^{src}} \mathbf{I}(i \in k)}{|\mathcal{C}_u^{src}|},  \text{$k=1,2,...,K$}
\end{equation}
Similarly, $u$'s interest distribution in the recommendation domain is defined as:
\begin{equation}
    p_{u}^{rec}(k) = \frac{\sum_{i \in \mathcal{C}_u^{rec}} \mathbf{I}(i \in k)}{|\mathcal{C}_u^{rec}|},  \text{$k=1,2,...,K$}
\end{equation}

After obtaining the interest distributions, we measure the degree of $u$'s interest shift by Jensen-Shannon (JS) Divergence of $p_{u}^{src}(k)$ and $p_{u}^{rec}(k)$:

\begin{align}
D^{J S}_{u}\left(p_{u}^{src} \| p_{u}^{rec}\right)&=\frac{1}{2} D_{K L}\left(p_{u}^{src} \| M \right)+\frac{1}{2} D_{K L}\left(p_{u}^{rec} \| M\right),\\
M& = \frac{p_{u}^{src}+p_{u}^{rec}}{2}.\nonumber
\end{align}
JS Divergence is bounded by 1, given the base 2 logarithm. If $ D^{J S}_{u} \geq 0.5$, we can assume $u$'s interest distribution shifts as the domain changes.

\begin{figure}[H] 
\centering 
\includegraphics[width=0.8\linewidth]{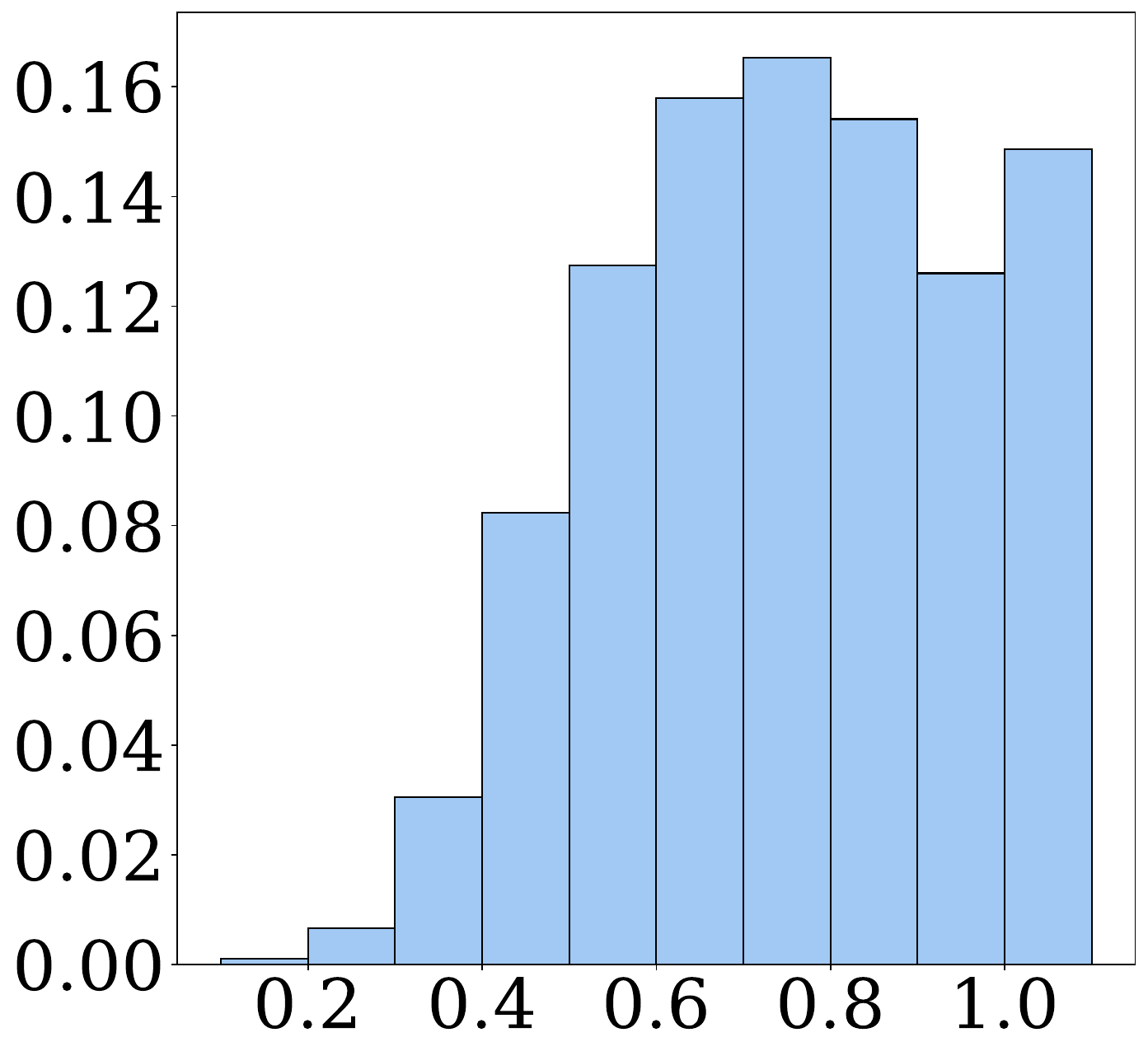} 
\caption{Histogram of JS Divergence Values of Selected User Interests.} 
\label{js_hist} 
\end{figure}

We randomly select $10000$ users and visualize their JS values by histogram. Figure \ref{js_hist} indicates that at least half of users have displayed unignorable changes in interest distribution across domains. Only 24.8\%  users have shown no obvious evidence of interest shift given $D^{J S}_{u} \leq 0.5$.

\subsubsection{Correlation between the search and the recommendation domains }
\label{domain_corr}
To see whether the information from the search domain would be an enhancement to the recommendation domain, we explore the correlation between the two domains. After randomly sampling $n=10000$ positive data points in the recommendation domain, we obtain: 
\begin{align}
\{(u^{1}, i^{1}, y^{1}, t^{1}),..,(u^{k}, i^{k}, y^{k}, t^{k}),... (u^{n}, i^{n}, y^{n}, t^{n})\},\end{align} where $y^{k} = 1$ for $k = 1,2,..,n$. 

$(u^{k}, i^{k}, y^{k}, t^{k})$ denotes that the user $k$ clicked the item $i^{k}$ in the recommendation domain at time $t^{k}$. $\vq_{t}^{k} = [q_1, q_2,...,q_m]$ is the list of queries input by the user $k$ before time $t^{k}$
 in the search domain. $\vs_{t}^{k}$ is the list of impression items resulting from each query in $\vq^{k}_t$. 

\begin{align}
\mathbf{s_{t}^{k}}= \{ \underbrace{i^{+}_{q_{1}1}, i^{+}_{q_{1}2},...,i^{-}_{q_{1}l_{1}}}_{\text{result items of } q_1},\underbrace{i^{+}_{q_{2}1}, i^{+}_{q_{2}2},...,i^{-}_{q_{2}l_{2}}}_{\text{result items of } q_2},..,\underbrace{i^{+}_{q_{m}1}, i^{+}_{q_{m}2},...,i^{-}_{q_{m}l_{m}}}_{\text{result items of } q_m}\}.
\end{align}
$i^{+}_{q_{m} 1}$ and $i^{-}_{q_{m} 2}$ denote items resulting from the query $q_m$, clicked and non-clicked, respectively. 

 We assume that the semantics behind a query can indicate a user's intention,  and then the impression items of this query are also reflecting the intention. Therefore, if a user's clicked items in the recommendation domain overlap with the items resulting from queries he inputs in the search domain, we can conclude that there is a correlation between his actions in the two domains. 

$i^k$ is the item clicked at time $t^{k}$ in the recommendation domain.
Then we calculate the following:
\begin{align}
\label{corr_f1}
R^{1}_{corr} = \frac{\frac{1}{n}\sum^{n}_{k=1}\mathbf{I}(i^{k} \in \mathbf{\vs^{k}_{t}})}{\frac{1}{n}\sum^{n}_{k=1}Pr(i^{k} \in \widehat{\bf{\vs^{k}_{t}}})},
\end{align}
where $\widehat{\bf{\vs^{k}_{t}}}$ denotes any list of items whose length is equal to $\mathbf{\vs^{k}_{t}}$'s length. $Pr(i^{k} \in \widehat{\bf{\vs^{k}_{t}}})$ is defined as
\begin{align}
Pr(i^{k} \in \widehat{\bf{\vs^{k}_{t}}}) = 1 - (1-\vp_{src}(i^{k}))^{\vert \mathbf{\vs^{k}_{t}} \vert},
\end{align}
 where $\vp_{src}(i^{k})$ is the probability of the item $i^{k}$ occurring in the search domain. If $ R^{1}_{corr} > 1 $, then it indicates that the actions in the search domain have implicit impacts on the actions in the recommendation domain.

We also consider the following score: 
\begin{align}
\label{corr_f3}
R^{2}_{corr} =\frac{1}{n}\sum^{n}_{k=1}\frac{\mathbf{I}(i^{k} \in \mathbf{\vs^{k}_{t}})}{Pr(i^{k} \in \widehat{\mathbf{\vs^{k}_{t}}})}.
\end{align}
If $R^{2}_{corr} >1$, there should be a correlation between the search domain and the recommendation domain.

In the implementation, we use $i^k$'s category to replace $i^k$. As shown in Table \ref{tab:correlation_R}, \textbf{KuaiSAR-small} and \textbf{KuaiSAR-large} are with $R^{1}_{corr} >1$ and $R^{2}_{corr} >1$.

\subsection{Experimental setup}

\subsubsection{Data processing}

 Following \cite{Kang2018SelfAttentiveSR, SESRec}, we group the interactions by users and sort them chronologically. In the recommendation domain, leave-one-out is the scheme to split datasets. After leave-one-out,  each user's most recent action is used for testing, with the second most recent action as the validation. The rest composes the train set. For each record with a timestamp $t$ in the recommendation domain, the user's historical queries until $t$ are made to be the search-aware enhancement features. To ensure each record has a search-aware enhancement, Only records of which the user simultaneously has search and recommendation histories are kept.
\subsubsection{Baselines}

In this study, we compare our model with the following baselines.

\begin{itemize}
\item \textbf{DIN} \cite{DIN2018} applies an attention module to capture user interests from historical behaviors with respect to the target item.
\item \textbf{PLE} \cite{Tang2020PLE} conducts a multi-task learning process using shared experts and task-specific experts with a progressive routing mechanism.
\item \textbf{DCN} \cite{wang2017deep} uses a cross-network to explicitly model feature interactions of bounded degrees, parallel to the feed-forward networks.
\item \textbf{SESRec} \cite{SESRec} employs distinct encoders to capture user interests from search and recommendation behaviors and disentangles similar and dissimilar representations between search and recommendation behaviors by contrastive learning, to boost recommendation performance.

\item \textbf{IV4Rec+} \cite{IV4REC} proposes a model-agnostic framework that uses search queries as instrumental variables to reconstruct user and item embeddings and enhance the performance in the recommendation domain in a causal learning manner. 
\end{itemize}



\subsubsection{Baseline adjustment}
The original \textbf{{SESRec}} \cite{SESRec} is designed for the sequential recommendation problem, and thus, we modify its dataset processing procedure and evaluation function to adapt it to a CTR prediction problem. One of the brilliant innovations in \textbf{IV4Rec+} \cite{IV4REC} is to leverage pre-trained text embeddings with dimension as 768, from Bert \cite{devlin2018bert} to complete the casual learning process. However, all texts in \textbf{KuaiSAR} are anonymized into integer numbers. Therefore, we replace pre-trained text embeddings with the learned embeddings from DIN \cite{DIN2018}. 

\begin{figure}[t]
    \centering
  \begin{subfigure}[t]{0.23\textwidth}  
\includegraphics[width=\textwidth]{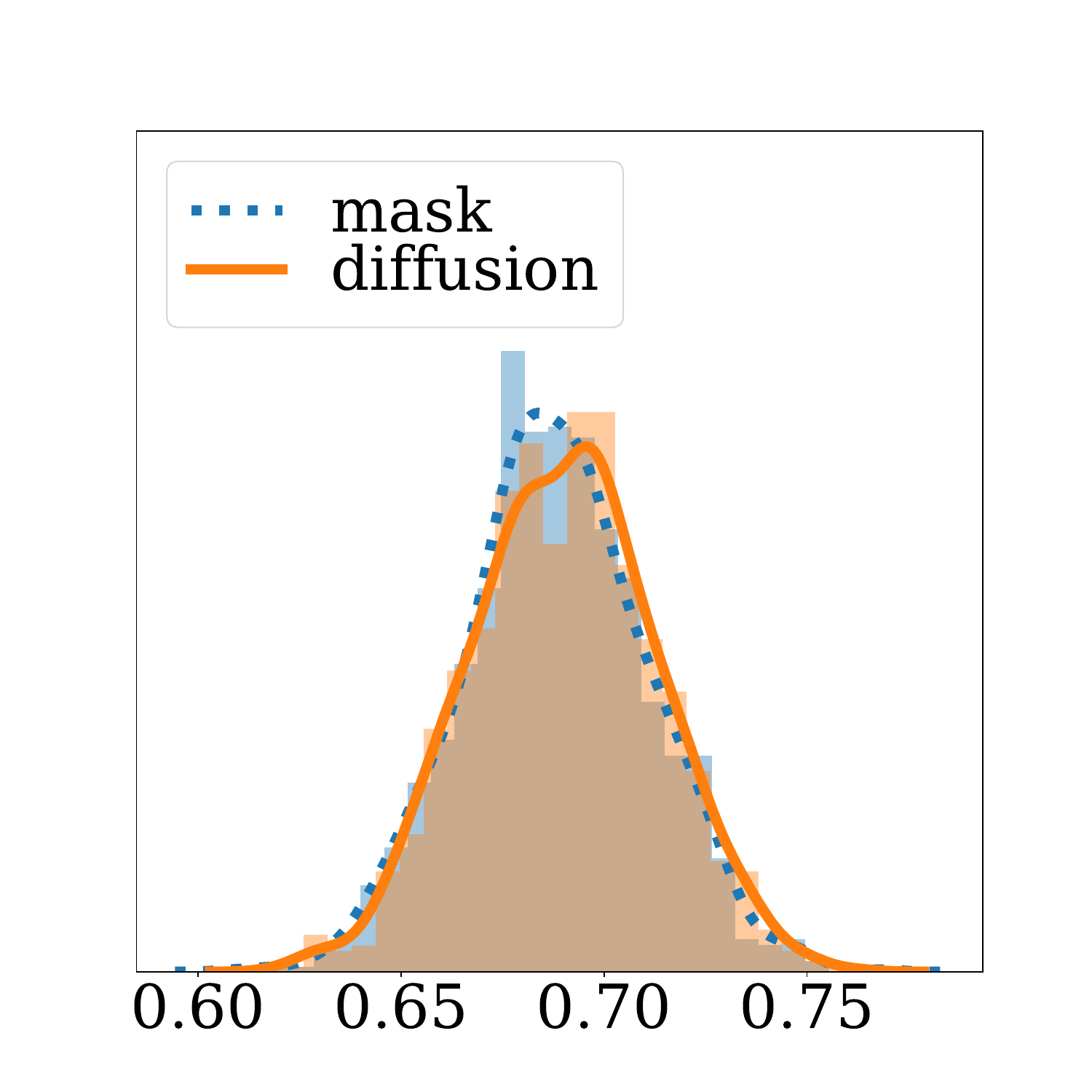}
\caption{Bootstrapped Mean Values}
  \end{subfigure}
    \hfill
\begin{subfigure}[t]{0.23\textwidth}  
\includegraphics[width=\textwidth]{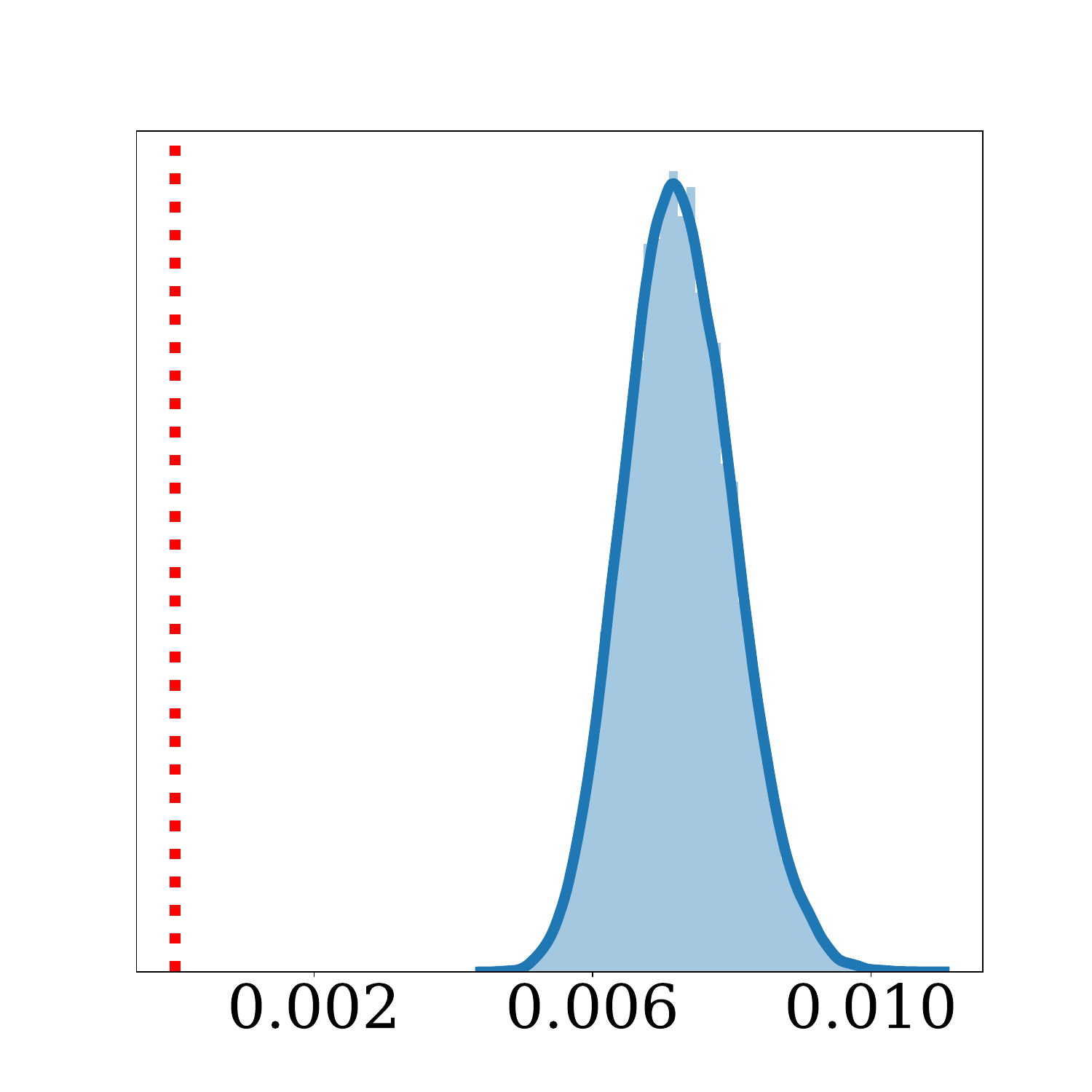}
\caption{Bootstrapped Difference in Sample Means. The red dotted line denotes the Null Hypothesis}
\end{subfigure}

    \caption{Statistic analysis on the performance of contrastive learning (CL) with Mask v.s. CL with diffusion-augmented positive samples.}
\label{case_fig1}
\end{figure}

\begin{figure}[t]
    \centering
  \begin{subfigure}[t]{0.23\textwidth}  
\includegraphics[width=\textwidth]{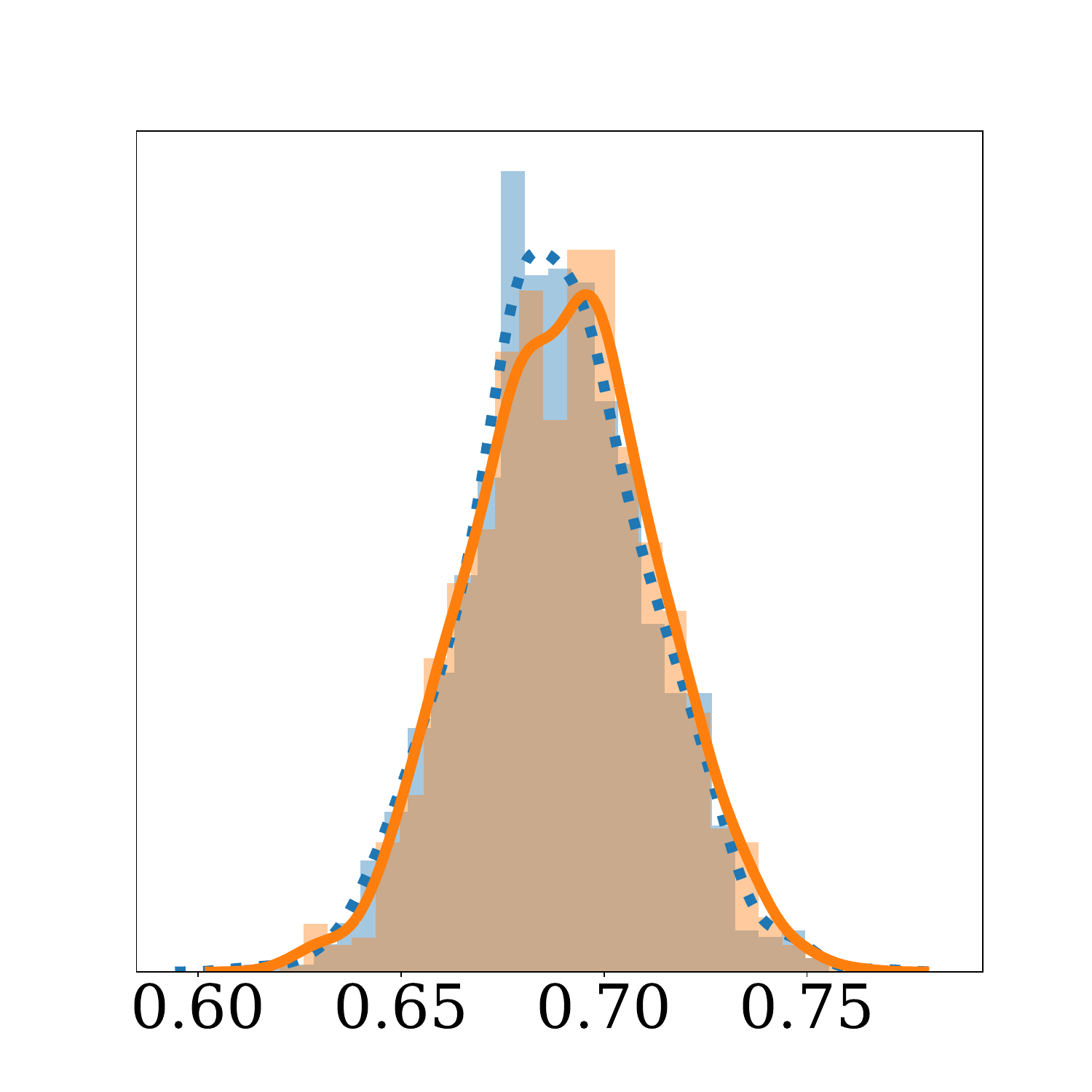}
\caption{Bootstrapped Mean Values. The dotted line is the one trained with $x^{1}_0$, and another is with $x^{2}_0$}
\label{case_fig2:a}

  \end{subfigure}
    \hfill
\begin{subfigure}[t]{0.23\textwidth}  
\includegraphics[width=\textwidth]{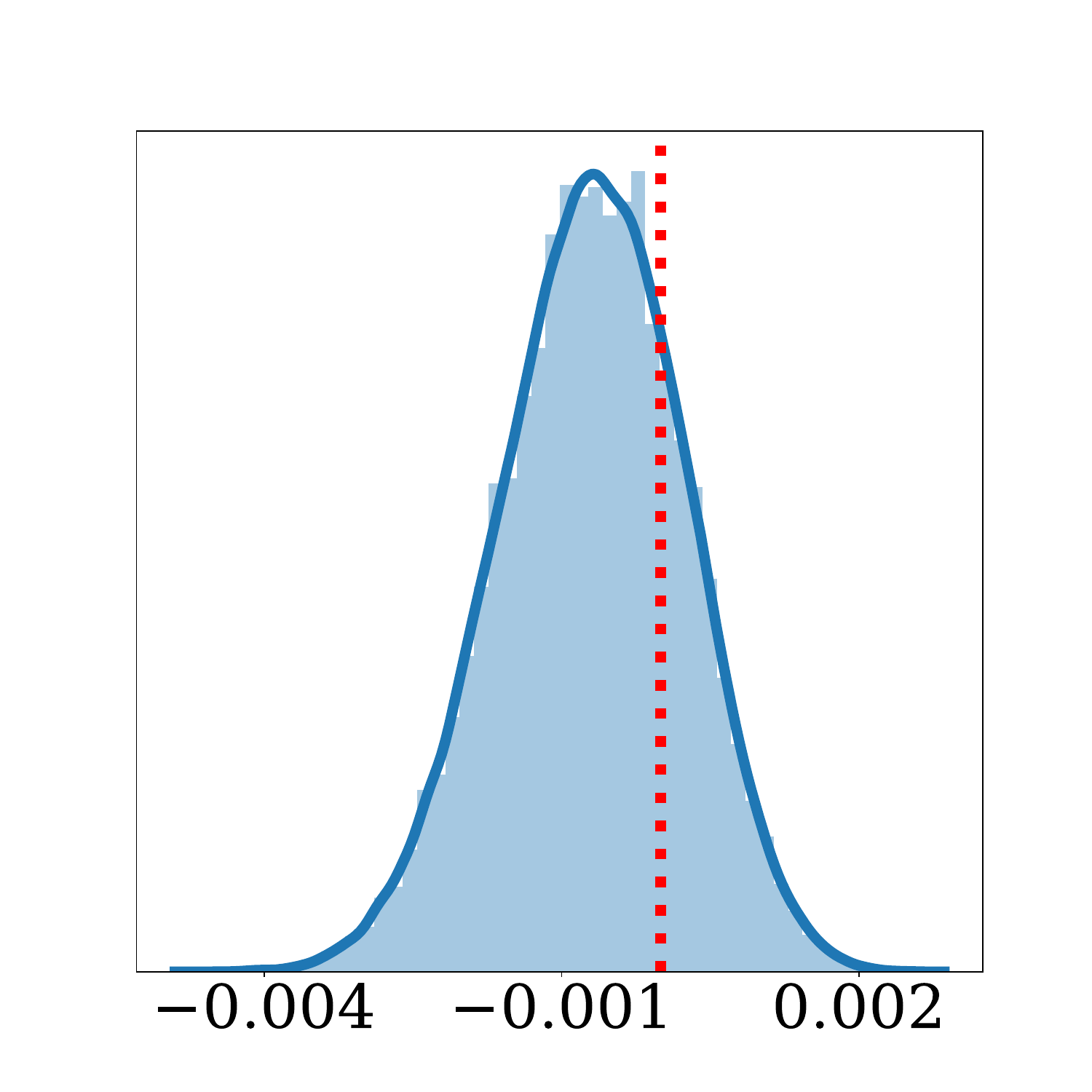}
\caption{Bootstrapped Difference in Sample Means. The red dotted line denotes the Null Hypothesis}
\label{case_fig2:b}

\end{subfigure}

    \caption{Statistic analysis on the impact of diffusion initialization values.}
\label{case_fig2}
\end{figure}

\subsection{Experimental result}
\begin{figure}[t]
    \centering
\includegraphics[width=0.23\textwidth]{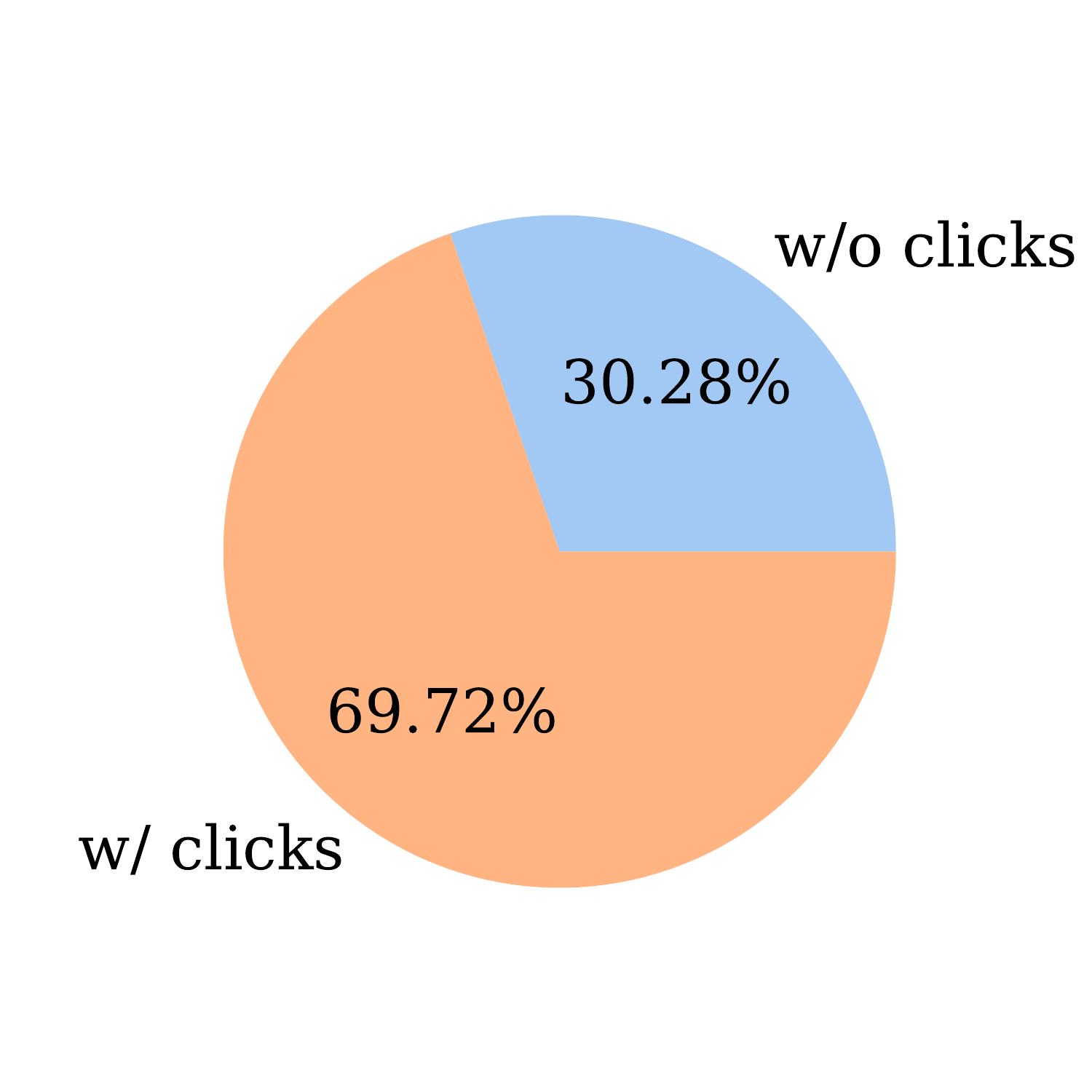}
\includegraphics[width=0.23\textwidth]{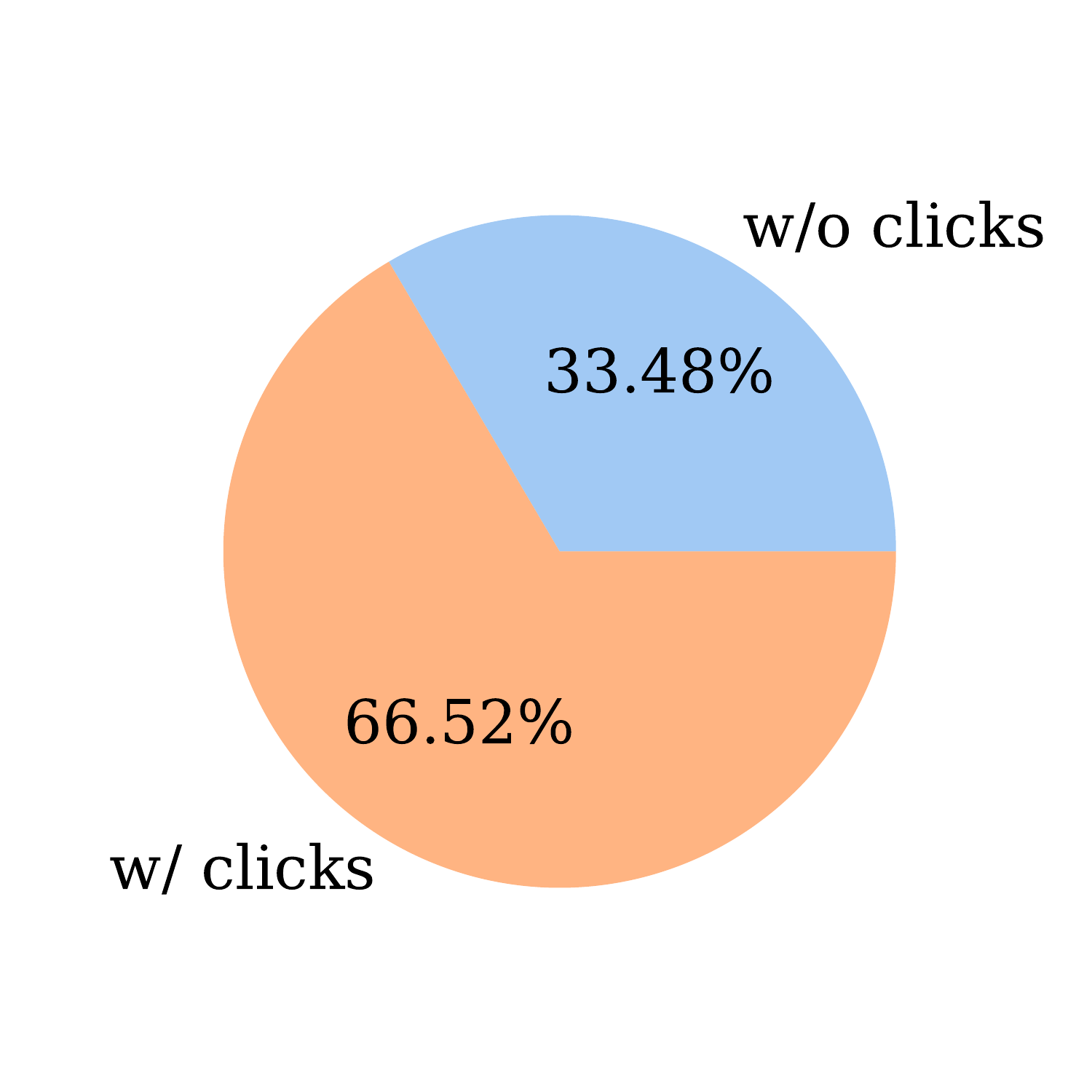}
    \caption{Data statistics. The comparison of queries with corresponding item clicks to queries with items clicked. Left: KuaiSAR-small; Right: KuaiSAR-large.}
\label{query_ratio}

\end{figure}
Table \ref{tab:main_label} reports the results on the three datasets. In QueryRec, $x_0$ is initialized by setting $r_p = 0.5$, $r_0 = 0$, $r_n = -0.5$ for diffusion augmentation. From the results, we can observe that: \begin{itemize}
    \item First, QueryRec outperforms all the baselines in KuaiSAR-small and KuaiSAR-large datasets. For the industrial dataset, it improves performance over commonly used models DCN and DIN. This implies the effectiveness of focus on the learning of queries, which brings meaningful information for capturing user's preferences
    \item SESRec performs the second best in KuaiSAR-small, while not performing well in KuaiSAR-large. This might indicate that the larger datasets require more effort to select between the transferable and non-transferable information. 
    \item IV4REC+ does not meet the expectations. The main reason is assumed to be the lack of real text embeddings. This reveals another limitation of IV4REC+ that it requires the dataset to have enriched text information. In comparison, our model is less restricted by the form of the data.
\end{itemize}
 
\begin{table}[t]
\caption{Comparisons of overall CTR prediction performance in recommendation domain of different methods on three datasets. AUC is used as the performance metric. The best model is highlighted in bold, respectively.}
\label{tab:main_label}
\resizebox{\columnwidth}{!}{
\begin{tabular}{c|cc|cc|cc}

\toprule
\textbf{Dataset} & \multicolumn{2}{c|}{\textbf{KuaiSAR-small}} & \multicolumn{2}{c|}{\textbf{KuaiSAR-large}} & \multicolumn{2}{c}{\textbf{Industrial}} \\ \hline
\textbf{Model}   & \textbf{AUC}         & \textbf{Imp \%}      & \textbf{AUC}         & \textbf{Imp \%}      & \textbf{AUC}       & \textbf{Imp \%}    \\ \hline
DIN              & 0.6793               & 0                    & 0.6892               & 0                    & 0.8943             & 0                  \\
DCN              & 0.6729               & -0.94                & 0.6749               & -2.1                 & 0.8948             & 0.02               \\
PLE              & 0.6825               & 0.47                 & 0.6879               & -0.19                & -                  & -                  \\
SESRec           & 0.6843               & 0.74                 & 0.6859               & -0.48                & -                  & -                  \\
IV4REC+          & 0.6640               & -2.25                & 0.6580               & -4.53                & -                  & -                  \\
QueryRec         & \textbf{0.6972}      & \textbf{2.64}        & \textbf{0.6996}      & \textbf{1.5}         & \textbf{0.8999}    & \textbf{0.63}      \\ \bottomrule
\end{tabular}}
\end{table}
\subsection{Ablation study}
We conducted an ablation study on the KuaiSAR-small dataset to examine the impact of each module in QueryRec as follows:
\begin{itemize}
    \item \textbf{DIN+query}: query is added as an extra user behavior feature for CTR model DIN.
    \item \textbf{DIN+NIP}: only the next item prediction is added to the framework
    \item \textbf{DIN+con}: contrastive learning is added without diffusion model to find positive samples for queries with no corresponding item click history. 
    \item \textbf{DIN+con w/diffusion}: contrastive learning is added with the diffusion model to find positive samples for queries with no corresponding item click history. 
\end{itemize}
Table \ref{ablation} displays the results of the ablation study. We could find out that:
\begin{itemize}
    \item Just incorporating query as an extra feature does not enhance the performance. This further confirms the assumption of the user interest shift. Information embedded in the query needs adaptation before employment in the recommendation domain.
    \item Each module shows positive improvements compared to DIN, while QueryRec provides the highest AUC value. This proves the effectiveness of each module of the framework.
    \item Compare DIN+con w/diffusion and DIN+con, it is obvious that augmenting positive data for queries with no item click history helps extract query-item relationship better.
    \end{itemize}
\begin{table}[t]
 \caption{Ablation study on KuaiSAR-small dataset. "w/" refers to "with", and "NIP" represents "Next Item Prediction".}
   \label{ablation}
    \resizebox{0.7\columnwidth}{!}{
\begin{tabular}{ccc}
\toprule
\textbf{Model}      & \textbf{AUC} & \textbf{Imp \%} \\ \hline
DIN                 & 0.6793       & 0               \\
DIN+query           & 0.6792       & -0.01           \\
DIN+NIP             & 0.6894       & 1.49            \\
DIN+con             & 0.6819       & 0.38            \\
DIN+con w/diffusion & 0.6890       & 1.43            \\
QueryRec            & 0.6972       & 2.64            \\ \bottomrule
\end{tabular}}
\end{table}
\subsection{Case study}

\label{case1}
 a common technique of common constrastive learning, if no positive items are available for a certain query, is to mask this query. However, as the proportion of such queries could not be neglected as indicated in figure \ref{query_ratio}, simply ignoring them might lead to poor learning of query-item relationship. Hence, we propose the diffusion augmentation to generate the positive samples for these queries. To justify our design, we do a comparison of contrastive learning (CL) with the masked sample and CL with the generated positive items by the diffusion model. We denote the AUC lists of these two as $AUC_1$ and $AUC_2$. Figure \ref{case_fig1} depicts the  bootstrapped distributions of mean values of the two AUC lists. 

Figure \ref{case_fig1} depicts the bootstrapped distribution of the differences between the means of $AUC_2$ and $AUC_1$, with the expectation of 0.7\%.  The calculated p-value $<$ $\alpha = 0.05$, with the null hypothesis that the means are equal between samples of two AUC lists. Based on these results, we can conclude that the AUC of CL with the diffusion would be higher than that of CL with the mask.

\subsection{Sensitivity study of different initialization values in diffusion process} 

\label{case2}
We compare two ways of the diffusion initialization, $x^{1}_0$ and $x^{2}_0$, so that we can get an insight into how sensitive the contrastive learning, and thus, the overall framework will be influenced by the choice of the initialization approach. In specific, $x^{1}_0$ is initialized by setting $r_p = 0.5$, $r_0 = 0$, $r_n = -0.5$, and  $x^{2}_0$ is initialized by setting  $r_p = 0.75$, $r_0 = 0$, $r_n = -0.75$. After the diffusion augmentation in \ref{augmentation}, we obtain two sets of positive samples from models trained with $x^{1}_0$ and $x^{1}_0$, respectively. The AUC list of CL with $x^{1}_0$ is $AUC_3$, and the one with $x^{2}_0$ is $AUC_4$. Similar to \ref{case1}, we compare $AUC_3$ and $AUC_4$ by the bootstrapping technique. From figure \ref{case_fig2}, there is no obvious difference of $A_3$ and $A_4$. Thus, our model is not sensitive to the difference between $x^{1}_0$ and $x^{2}_0$. In other words, the model is robust with different implementations of diffusion augmentation.

\section{Conclusion}
In this paper, we have introduced a novel method for search-enhanced CTR prediction for the recommendation domain. Our method centers on developing an informative query representation, which is seamlessly incorporated into the CTR prediction model. This integration facilitates the transfer of knowledge from the search to the recommendation domain. The query representation is refined using a next-item prediction module that aligns the query more closely with user interests in the recommendation context. Additionally, our use of diffusion-augmented contrastive learning enhances the extraction of query-item relationships. The efficacy of our framework is proved by results from three datasets and an ablation study, which collectively validate the effectiveness of each component.

\bibliographystyle{ACM-Reference-Format}
\balance
\bibliography{main}

\end{document}